\documentclass{pasj00}
\usepackage{graphicx}
\usepackage{epsfig}
\usepackage{lscape}

\begin{document}

\SetRunningHead{Author(s) in page-head}{Running Head}
\Received{2006/07/21}%{yyyy/mm/dd}
\Accepted{2006/08/24}%{yyyy/mm/dd}

\title{Do Ultraluminous X-Ray Sources  Really Contain Intermediate-mass Black Holes?}

 \author{%
   Kiki \textsc{Vierdayanti}\altaffilmark{1},
   Shin \textsc{Mineshige}\altaffilmark{1},
   Ken \textsc{Ebisawa}\altaffilmark{2}
  and
   Toshihiro \textsc{Kawaguchi}\altaffilmark{3} 
}
 \altaffiltext{1}{Yukawa Institute for Theoretical Physics, Kyoto University, Sakyo-ku, Kyoto 606-8502}
 \altaffiltext{2}{Institute of Space and Astronautical Science, Japan Aerospace Exploration Agency (ISAS/JAXA), \\
Yoshinodai 3-1-1, Sagamihara,
Kanagawa, 2298510, Japan}
 \altaffiltext{3}{Department of Physics and Mathematics, Aoyama Gakuin University, Kanagawa, 229-8558}
 \email{kiki@yukawa.kyoto-u.ac.jp}

%% `\KeyWords{}' always has to be placed before `\maketitle'.
\KeyWords{accretion, accretion disks --- black hole physics --- X-rays: individuals (NGC 5204 X-1, NGC 4559 X-7, NGC 4559 X-10, NGC 1313 X-2) --- X-rays: stars} %Do NOT move this preamble from here!

\maketitle

\begin{abstract}
An open question remains whether Ultraluminous X-ray Sources (ULXs) 
really contain intermediate-mass 
%($\sim$ several hundreds $M_\odot$) 
black holes (IMBHs). 
We carefully investigated the XMM-Newton EPIC spectra of the four ULXs
that were claimed to be strong candidates of IMBHs 
by several authors.
We first tried fitting by the standard spectral model of 
disk blackbody (DBB) + power-law (PL), 
finding good fits to all of the data 
%with low innermost temperatures of the disk
%($0.1 < kT_{\rm in} < 0.5$ keV)
%and extremely large innermost radii. 
%These apparently support 
%the IMBH interpretation of ULXs 
, in agreement with others.
We, however, found that the PL component dominates  the DBB component
at  
%almost entire energy ranges in 
$\sim$ 0.3 to 10 keV.  Thus, the black hole parameters
 derived
solely from the minor DBB component are questionable.
Next, we tried to fit the same data by the ``$p$-free disk model'' without the PL component, 
assuming the 
effective temperature profile of $T_{\rm eff} \propto r^{-p}$ where $r$ is the disk radius. 
Interestingly, 
in spite of one less free model parameters,
we obtained similarly good fits with much
higher innermost disk temperatures, 
$1.8 < kT_{\rm in} < 3.2$ keV.
More importantly, we obtained $p \sim 0.5$, just the value predicted 
by the slim (super-critical) disk theory, rather than $p = 0.75$  that is
expected from the standard disk model.
The estimated black hole masses from the $p$-free disk model
 are much smaller; $M\ltsim 40 M_\odot$.
% ,and the Eddington ratios are larger; $L/L_{\rm E} \gtsim 0.4$.
Furthermore, we applied a
more sophisticated slim disk model by Kawaguchi (2003, ApJ, 593,69), and obtained good fits
with roughly consistent black hole masses.
We thus conclude that the central engines of these ULXs are super-critical 
accretion flows to stellar-mass black holes. 
%We suggest that all the previous fitting results of ULXs with 
%the DBB+PL model resulting in very low disk temperatures should be re-examined, 
%since these data are very likely to be fitted as well with 
%the high temperature slim disk ($p$-free) 
%model, which is physically sounder,
%and to strengthen our ULX model  of the super-critical accretion flow to
%stellar-mass black holes. 
\end{abstract}

\section{Introduction}
X-ray data analysis has become one of the most widely studied subjects in these days, %since the launch of Einstein satellite in 1978
and has made great contributions in the progress of astrophysics.
This is particularly true in the field of astrophysical black holes.
Thanks to rapid improvement of the sensitivity of X-ray detectors,
new types of X-ray sources have been discovered.
Some of them seem to be black holes which were not known before.

Recent X-ray observations on many nearby spiral galaxies showed that 
the X-ray radiation consists of emission from several discrete X-ray sources, such as
hot gaseous media, active galactic nucleus (AGN) 
and the point-like off-center X-ray sources whose X-ray luminosities significantly 
exceed the Eddington luminosity of a neutron star (\cite{15}). 
The bright point-like off-center X-ray sources which cannot be identified with young supernova remnants are known as ultraluminous X-ray sources (ULXs; e.g., \cite{25}; 
\cite{34}). 
The ULXs, despite the fact of being very luminous ($L_{\rm x}\sim 10^{39 - 41}{\rm erg} {\rm s}^{-1}$), 
cannot be explained as collections of many sources with each luminosity 
less than the Eddington limit,
 since many of the ULXs exhibit significant time variabilities (e.g., \cite{25}). Therefore, one reasonable assumption is
 that the ULXs are single compact objects powered by accretion flow. 
Then, how can we understand the extremely
large luminosities exceeding the Eddington luminosity for a mass of 
$\sim $10 $M_\odot$?

A possibility that ULXs are only bright toward our line of sights (beaming model) was
proposed  (e.g. \cite{22}, \cite{23}, \cite{35}),
but is not generally accepted now. In fact, 
Wang (2002) has shown 
the existence of a bright nebula surrounding the ULX M81 X-9,  and some ULXs also exhibit extended optical nebulae (\cite{51}). There is an evidence that some
ULXs are associated with a spreading wave of star formation,  which supports the idea that 
many ULXs, in particular those located in star forming galaxies, are high mass X-ray binaries (\cite{21}). 
These facts indicate that the ULX surrounding nebulae are  powered by the central high mass X-ray sources,  and that strong X-ray beaming toward our direction is  unlikely.

Then, the question is, how massive the central compact objects  are.
At present, there are basically two distinct lines of thoughts:
sub-Eddington accretion onto an intermediate 
(several hundreds $M_\odot$)
mass black hole (IMBH), and super-Eddington (super-critical) accretion onto a stellar-mass black hole.
The success in fitting several ULXs spectra with a multicolor disk blackbody (DBB) and a power-law (PL) model  supports the idea of IMBHs for the ULXs 
(\cite{29}, \cite{30}; \cite{9}, \cite{36}), since
the cool inner disk temperatures and  the large innermost radii, obtained through model fitting, 
suggests a black hole mass within the IMBH range.
The other line of thoughts, 
 supporting a super-Eddington accretion flow, is the slim disk model with
stellar mass black holes (\cite{48}; \cite{20}, \cite{14}; \cite{33}).
The present article concerns with the super-Eddington thick accretion
disks introduced in the early 1980s by the Warsaw group and their
collaborators (e.g. \cite{1}; \cite{2}) and with the slim accretion disks, introduced 
in
the late 1980s by the Warsaw and the Kyoto groups (\cite{3},
 \cite{4}). In these papers, the very
possibility of the super-Eddington beaming in the thick accretion disk
funnels, crucial for arguments presented here, was explicitly 
recognized,
discussed and stressed.

The IMBH notion faces a serious problem that the formation of IMBH itself remains unsettled (e.g., \cite{52}).
In contrast, the
 alternative possibility of the super-critical accretion has been poorly investigated so far due
mainly to our limited knowledge about properties of the super-critical flow.
The situation has been remarkably improved in the past few years,
however, since the basic tools for investigating high-luminosity accreting systems
are now available.
In the present paper, we investigate the XMM-Newton EPIC spectra of the four ULXs
which had been claimed as strong IMBH candidates, using theoretical
 super-critical models to see whether the 
super-critical accretion takes place in these ULXs or not.
Our final goal is
to determine the mass of the central black holes
 and settle down the issue regarding the origin of ULXs.

We applied accretion disk spectral models to the observed
ULX energy spectra, constrain the innermost disk radius, and estimate
the black hole mass. A very similar method as the one used in the
present paper was successfully applied to Galactic
black hole binaries (e.g., \cite{12}; \cite{13}; \cite{17}). 
Several authors have recently done elaborate mass and angular momentum
estimates by spectral fitting for microquasars, i.e. by a very similar
method as the one used in the present paper (\cite{38}; \cite{27}; \cite{10}). In the present study, however, we focus on the mass estimation and do not attempt to estimate the spin of the black hole.

The mass of the ULXs can also be accurately estimated by a different
method than that described in our article. For this other independent
estimate, double peak QPOs in the 3:2 resonance should be detected in 
ULXs
(\cite{5}). This is because these QPOs scale inversely 
with
the mass of the source, which was first shown for microquasars by
McClintock and Remillard (2003), and more recently for low-mass Seyfert
galaxies by Lachowicz et al. (2006). There is also an
indication that the same scaling is true for a (possible) detection of
double peak 3:2 QPOs in SgrA* (\cite{41}), while for the ULXs, in general, it has not been detected yet. According to Mucciarelli et al. (2006), the only ULX where a QPO has been discovered at present is M82 X-1.
Scaling the frequency inversely to the BH mass, the observed QPO 
frequency
range (from a few tens to a few hundreds mHz) would yield a black hole mass
anywhere in the interval from a few tens to a few thousands solar mass
(\cite{40}; \cite{32}).
 
The Plan of this paper is as follows:
We first describe the X-ray data which we used in the present study
and our methods of fitting in Section 2.
In Section 3, we then give the results of fitting, together with their implications
on the black hole mass and mass accretion rates.
Section 4 is devoted to discussion, and the final Section 5 concludes the paper.

\section{Data and Fitting Methods}

\subsection{Data}

We perform spectral analysis of four ULXs by using XMM-Newton observational data. The data were extracted using the XMM-Newton SAS version 6.5.0 tools. Detail extraction process will be explained separately for each object. Response files were made using the SAS tools rmfgen and arfgen for all the data. 

\subsubsection{NGC 5204 X-1}
NGC 5204 X-1 is located $\sim 0.3$ kpc from the center of a nearby, $D=4.8$ Mpc, Magellanic-type galaxy (\cite{37}). The typical X-ray luminosity of this source is of the order of $2 \sim 6 \times 10^{39} {\rm erg}\ {\rm s}^{-1}$ (0.5 -- 8 keV) (\cite{36}).

NGC 5204 X-1 data for our analysis were obtained on 2003 January 6 (observation ID 0142770101). Following Roberts et al. (2005), we set flag = 0 and pattern $\leq 4$ for pn data , while \#XMMEA\_EM flag and pattern $\leq 12$ were used for the MOS data. No time filters were applied to the data since the background is less than 10 count ${\rm s}^{-1}$ in the pn detector. Source spectra and light curves were extracted in $36-{\rm arcsec}$ radius circle centered on the source and $44-{\rm arcsec}$ radius region nearby was used to produce background data.

\subsubsection{NGC 4559 X-7 and X-10}
NGC 4559 X-7 and X-10 are two brightest sources in NGC 4559 ($D=9.69$ Mpc). X-7 was located at the outer spiral arms, while X-10 was about 0.3 kpc from the optical nucleus of NGC 4559. They are also known as X1 and X4 respectively in Roberts and Warwick (2000), and IXO65 and IXO66 in Colbert and Ptak (2002) respectively (\cite{9}). 

The data were obtained on 2003 May 27 (observation ID 0152170501) for both X-7 and X-10. 
However, X-10 was only seen in EPIC-pn since the EPIC-MOS cameras were operated in small window mode. 
Following Cropper et al. (2004), we used $30-{\rm arcsec}$ radius centered to the source for X-7, while for X-10 we used $20-{\rm arcsec}$ radius with an exclusion of $6-{\rm arcsec}$ radius centered on the nucleus of the galaxy ($12^{h} 35^{m} 57.^{s}64$; $+27^{\circ} 57' 35''.8$, J2000.0). 
Source free region on the same CCD was selected as background for the EPIC-pn data while region as close as possible to the source were selected for EPIC-MOS X-7 data. 
The MOS data were filtered using pattern $\leq 12$ and \#XMMEA\_EM flag, while the pn data were filtered using pattern $\leq 4$ and \#XMMEA\_EP flag.

\subsubsection{NGC 1313 X-2}

NGC 1313 X-2 is one example of ULX sources in a nearby spiral galaxy, NGC 1313 ($D=3.7$ Mpc).
This ULX is located at approximately 8 kpc from the photometric center of the galaxy (\cite{7}).
The data for our analysis were obtained on 2000 October 17 (observation ID 0106860101).

We performed the standard procedures but somehow we failed to create response files for the MOS data, leaving us only the pn data to be analyzed.
We set flag = 0 and pattern $\leq 4$ for the pn-data.
Following Miller et al. (2003), we used 24-arcsec radius centered to the source to obtain source spectra, while background counts were extracted in an annulus between 24-arcsec and 30-arcsec.

\subsection{Fitting Methods}
The source spectra were grouped to a minimum of 20 count/bin before fitting using XSPEC version 11.3. We first try the standard spectral model of disk blackbody (DBB) + power-law (PL). We use the results as comparison with other papers. Next, we tried to fit the data by the ``$p$-free disk model'', assuming 
the effective temperature profile of $T_{\rm eff} \propto r^{-p}$,
where $r$ is the disk radius
 (\cite{31}). 
The $p$-free model is a potentially useful spectral model, since despite its simpleness it is very powerful 
for discriminating slim disk with $p\sim 0.5$ from the standard disk with $p = 0.75$ (\cite{46}).

We also tried to fit the data with a more sophisticated slim disk model 
calculated by Kawaguchi (2003),
who calculated more realistic emergent spectra of super-critical flow
based on the slim disk model, taking into account relativistic and Compton-scattering effects.
We will show the results separately apart from those two mentioned above, to avoid muddle, and we will give a brief comment on these results. 

The effect of Galactic absorption was taken into account by using wabs model (in XSPEC), where we fixed the value at $1.5\times10^{20} {\rm cm}^{-2}$ for NGC 5204 X-1 and NGC 4559 X-7 and X-10, and $3.9\times10^{20} {\rm cm}^{-2}$ for NGC 1313 X-2 from Dickey and Lockman (1990).
The effect of absorption external to our galaxy was also considered and again we used wabs model but we let this value free throughout the fitting.

\section{Spectral Analysis}
The results of spectral analysis for each ULX investigated in this paper will be presented separately 
for three distinct spectral models: DBB+PL models (in sections 3.1 and 3.2),
$p$-free models (in section 3.3), and Kawaguchi's models (in section 3.4).
In addition, we summarize the fitting results in Table 1. % for the readers' convenience.

\subsection{DBB+PL model}
We first present our results of fitting based on the conventional approach;
i.e., by using the DBB+PL model.
As will be explicitly demonstrated below,
we will basically confirm the previous results supporting the IMBH hypothesis.

\subsubsection{NGC 5204 X-1}
We fit the MOS data over 0.3 -- 8 keV and the pn data over 0.3 -- 10 keV. We obtained a good fit to the data with DBB+PL model (see Fig.1.). The reduced chi-squared is $1.05$ with the inner disk temperature, $kT_{\rm in}=0.25\pm0.03$ keV, and photon index, $\Gamma = 1.92\pm0.06$. The low temperature obtained from the fitting apparently supports the IMBH interpretation of ULXs, in agreement with Roberts et al. (2005). The flux obtained from the fitting by using this model is, $f_{\rm x}$ (0.3 -- 10 keV)$=1.66\times10^{-12} {\rm erg}\ {\rm cm}^{-2}{\rm s}^{-1}$. 
These are in reasonable agreement with the previous results by Roberts et al. (2005);
that is,
$kT_{\rm in}=0.21\pm0.03$ keV, and $\Gamma = 1.97\pm0.07$.

\subsubsection{NGC 4559 X-7 and X-10}
We next fit both the MOS and the pn data for NGC 4559 X-7 and the pn data for NGC 4559 X-10 over 0.3 -- 10 keV. 
As before, we fit the data with the DBB+PL model and we obtained reduced chi-squared of 0.96 and 0.92 for X-7 and X-10, respectively.
The inner disk temperature, $kT_{\rm in} =0.17 \pm 0.09$ keV, and $\Gamma = 2.17 \pm 0.05$, were obtained for NGC 4559 X-7, while $kT_{\rm in} = 0.48 \pm 0.25$ keV and $\Gamma = 1.94 \pm 0.13$ were obtained for NGC 4559 X-10. 
However, for the case of X-10, we could actually obtained a good fit by the PL model alone, giving reduced chi-squared of 0.92 and $\Gamma = 1.99 \pm 0.03$. 
%The flux is $8.36\times10^{-13} {\rm erg}\ {\rm cm}^{-2} {\rm s}^{-1}$ 
% and $9.42\times10^{-13} {\rm erg}\ {\rm cm}^{-2} {\rm s}^{-1}$ for NGC 4559 X-7 and X-10, respectively.

For reference, the previous fits by Cropper (2004) obtained
$kT_{\rm in}=0.148\pm0.006$ keV, and $\Gamma = 2.23\pm0.05$ for X-7,
in good agreement with our results.
As for X-10, Cropper et al. (2004) only shows the fitting result with single component power-law. 

\subsubsection{NGC1313 X-2}
We fit the pn data of NGC 1313 X-2 over 0.2 -- 10 keV with the DBB+PL model and found a good fit with reduced chi-squared of 0.96.
As in previous cases, we obtain low inner disk temperature, $kT_{\rm in} =0.27 \pm 0.04$ keV, and $\Gamma = 2.01 \pm 0.13$.

For comparison, Miller et al. (2003) found
$kT_{\rm in}=0.16\pm0.16$ keV, and $\Gamma = 2.3\pm0.2$, for NGC 1313 X-2,
from fitting the MOS data with the DBB+PL model.
The difference of the detectors is likely the origin of the small discrepancies.

%%%%%%%%%%%%%%%%%%%%%%%%%%%%%%%%%%%%%%%%%%%%%%%%%%%%%
\begin{figure*}
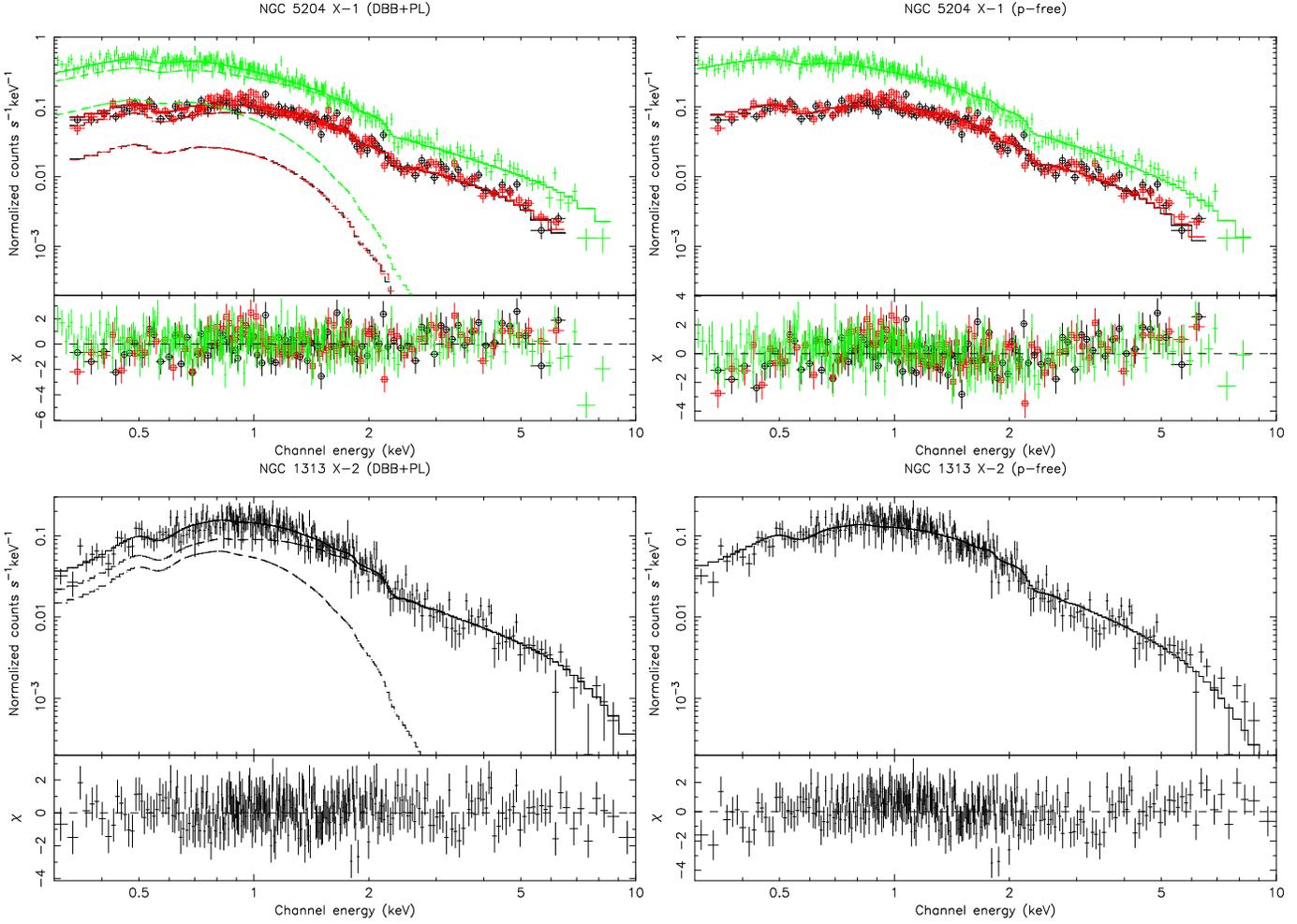

  \begin{center}
\centerline{\epsfig{file=pasj5204x1stdcolor.ps,width=6.5cm,angle=270}\epsfig{file=pasj5204x1pfreecolor.ps,width=6.5cm,angle=270}}
\centerline{\epsfig{file=pasj1313x2stdcolor.ps,width=6.5cm,angle=270}\epsfig{file=pasj1313x2pfreecolor.ps,width=6.5cm,angle=270}}
    %%% \FigureFile(width,height){filename}
  \end{center}
  \caption{Best fitting spectra and the fit residuals with the DBB+PL model (left panels) and
 p-free disk model (right panels) for NGC 5204 X-1, NGC1313 X-2, NGC 4559 X-7 and X-10. 
Open circles, open squares and dots represent mos1, mos2, and pn data, respectively. Dashed lines in the left panels represent spectral components.}\label{Figure:1}
\end{figure*}
\addtocounter{figure}{-1}
\begin{figure*}
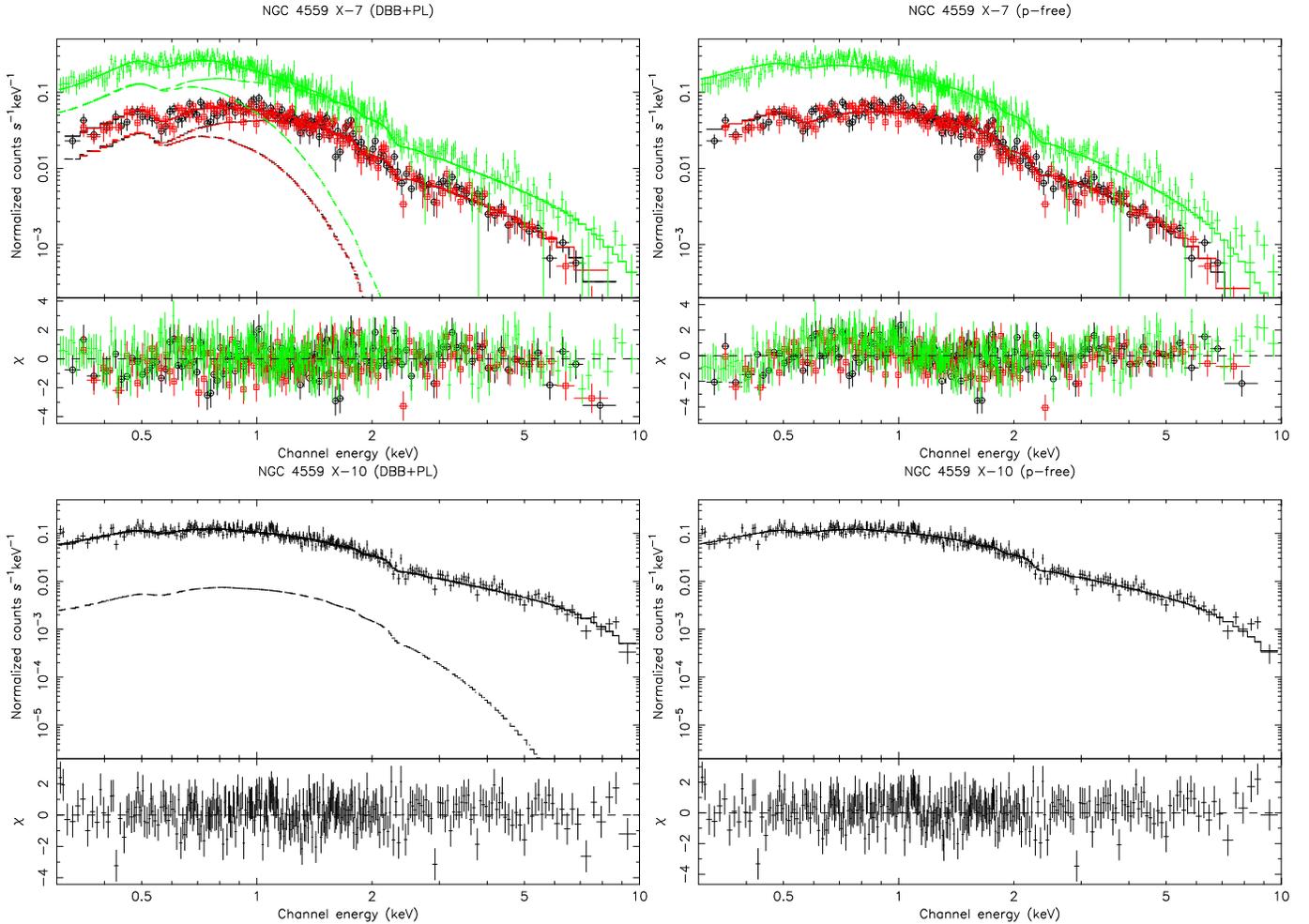

  \begin{center}
\centerline{\epsfig{file=pasj4559x7stdcolor.ps,width=6.5cm,angle=270}\epsfig{file=pasj4559x7pfreecolor.ps,width=6.5cm,angle=270}}
\centerline{\epsfig{file=pasj4559x10stdcolor.ps,width=6.5cm,angle=270}\epsfig{file=pasj4559x10pfreecolor.ps,width=6.5cm,angle=270}}
  \end{center}
  \caption{{\em Continued.}}\label{Figure:1}
\end{figure*}

%%%%%%%%%%%%%%%%%%%%%%%%%%%%%%%%%%%%%%%%%%%%%%%%%%%%%

\subsection{Interpretation based on DBB+PL model}
\subsubsection{Basic methodology to derive black-hole mass}
When we obtain a good fit to the observed spectra by the DBB model,
we can easily estimate the black-hole mass and the Eddington ratio,
$L/L_{\rm E}$ (with $L$ and $L_{\rm E}$ being disk luminosity
and the Eddington luminosity, respectively), based on the standard disk theory.
The basic methodology is summarized in Makishima et al. (2000).

Following Makishima et al.(2000), the bolometric luminosity of an optically thick accretion disk can be written as
 \begin{equation}
  L_{\rm bol} = 2 \pi D^2 f_{\rm bol} (\cos{i})^{-1}
 \label{1}
 \end{equation}
with {\it i} is the inclination of the disk (${\it i}=0$ corresponds to face-on geometry) and {\it D} is the distance.
This $L_{\rm bol}$ is related to the maximum disk color temperature $T_{\rm in}$ and the innermost disk radius $R_{\rm in}$ as

 \begin{equation}
  L_{\rm bol} = 4 \pi (R_{\rm in}/\xi)^2 \sigma (T_{\rm in}/\kappa)^4
 \label{2}
 \end{equation} 
where $\kappa \sim 1.7$ (\cite{39}) is the ratio of the color temperature to the effective temperature, or spectral hardening factor, and $\xi = 0.412$ is correction factor reflecting the fact that $T_{\rm in}$ occurs at somewhat larger than $R_{\rm in}$.
Hence, the innermost disk radius is 

 \begin{equation}
  R_{\rm in} = \xi \kappa^2 \sqrt {\frac{L_{\rm bol}}{4\pi\sigma T_{\rm in}^4}}
 \label{3}
 \end{equation}
We may identify $R_{\rm in}$ with the radius of the last stable Keplerian orbit. Thus, we may in general write

 \begin{equation}
  R_{\rm in} =3\beta R_{\rm s} = 8.86\beta (\frac{M}{M_{\odot}}) {\rm km}
 \label{4}
 \end{equation} 
where $R_{\rm s}$ is the Schwarzschild radius ($=2GM/c^2$), by which we can determine the black hole mass. Note that $1/6\leq \beta\leq1$, $\beta=1/6$ for extremely rotating Kerr black hole and $\beta=1$ for Schwarzschild black hole.

Here, we use 0.3 -- 10 keV flux, $f_{\rm [0.3 - 10 keV]}$, from which we derive 0.3 -- 10 keV luminosity, $L_{\rm [0.3 - 10 keV]}$. We assume $L_{\rm bol}\approx L_{\rm [0.3 - 10 keV]}$ because all significant energy range from 0.3 to 10 keV. Therefore, using $T_{\rm in}$ and $f_{\rm [0.3 - 10 keV]}$ from the fitting results, we can determine the value of $R_{\rm in}$, black hole mass, and also $L_{\rm [0.3 - 10 keV]}/L_{\rm E}$, where $L_{\rm E}$ is the Eddington luminosity ($=1.5\times10^{38}(M/M_{\odot}) {\rm erg}\ {\rm s}^{-1}$).

\subsubsection{Problem of DBB+PL model fitting}
By using the equations above and the values of $T_{\rm in}$ and $f_{\rm [0.3 - 10 keV]}$ from the fitting results, we obtained, for the case of NGC 5204 X-1, $R_{\rm in}=2.54\times10^3(\cos i)^{-1/2}$ km and the derived black hole mass of $287 \beta^{-1} (\cos i)^{-1/2}M_{\odot}$. We also obtained $L_{\rm [0.3 - 10 keV]}/L_{\rm E} = 0.05\beta (\cos i)^{-1/2}$. 
In the same way, we obtained relatively large $R_{\rm in}$ and hence large black hole mass exceeding 100 $M_\odot$ for other sources (see Table 2).

%In the same way, we obtained $R_{\rm in}=7.88\times10^3(\cos i)^{-1/2}$ km, the derived black hole mass of $889\beta^{-1} (\cos i)^{-1/2}M_{\odot}$. and $L_{\rm bol}/L_{\rm E} = 0.03 \beta (\cos i)^{-1/2}$ for NGC 4559 X-7 and $R_{\rm in}=1.05\times10^3(\cos i)^{-1/2}$ km, the derived black hole mass of $118 \beta^{-1}(\cos i)^{-1/2}M_{\odot}$. and $L_{\rm bol}/L_{\rm E} = 0.3 \beta (\cos i)^{-1/2}$ for X-10. 

To summarize, 
we found through the fitting with the DBB+PL model
that the derived black hole masses by far exceed 100 $M_\odot$ and are, hence, 
within the IMBH regime. Thus, fitting the data with the conventional model supports the idea of sub-critical accretion onto the IMBHs. 

However, we here point out a serious problem in this interpretation.
That is, the spectral decomposition 
shows that the power-law component predominates in all energy region in question
(see Fig.2 for the case of NGC 5204 X-1). In other words, the disk component only gives  a minor contribution in the entire energy ranges 
for all the sources.
Therefore, one is never allowed to use equations (1)--(4), which are derived 
on the assumption that the radiation is for 100 \% from the DBB component.
The black hole mass values derived above cannot be so reliable.

%%%%%%%%%%%%%%%%%%%%%%%%%%%%%%%%%%%%%%%%%%%%%%%%%%%%%
\begin{figure*}
  \begin{center}
\centerline{\epsfig{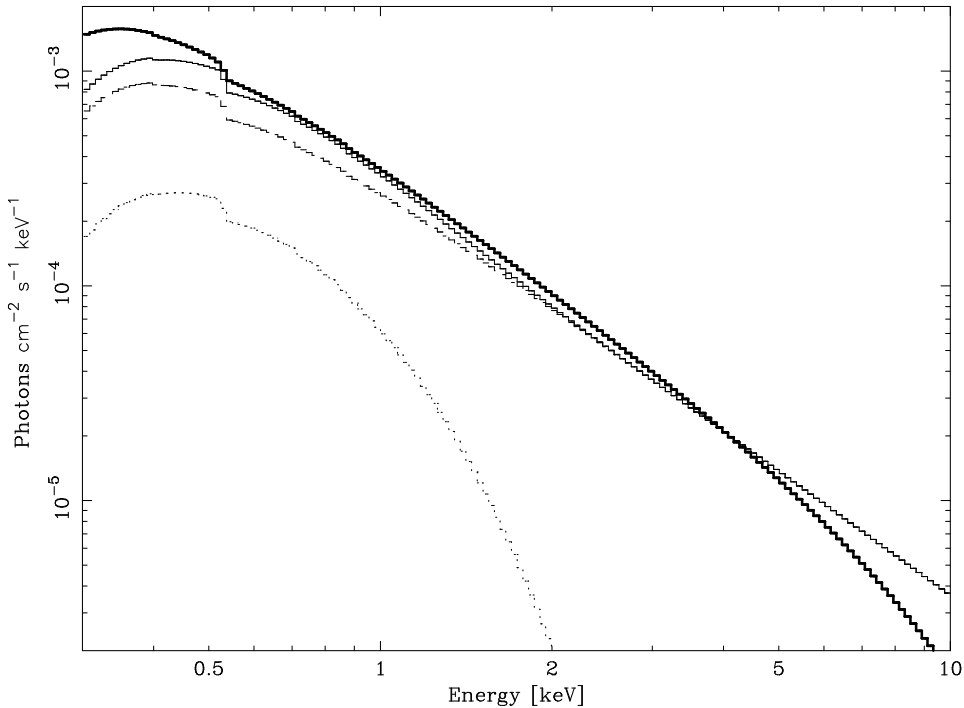}}
  \end{center}
  \caption{Spectral decompositions of the fits for NGC 5204 X-1. The thinner lines represent, from bottom to top, disk component, power-law component, and total component, while the thicker line represents p-free model}\label{Figure:2}
\end{figure*}

%%%%%%%%%%%%%%%%%%%%%%%%%%%%%%%%%%%%%%%%%%%%%%%%%%%%

\subsection{$p$-free model}
Next, we tried to fit the same data set by the simplified super-critical
disk model alone; i.e., the $p$-free disk model. 

\subsubsection{Fitting results}
Firstly, we tried the case of NGC 5204 X-1. Surprisingly we also obtained a good fit with
this model, although only a single component is considered (see Fig.3).
The reduced chi-squared is 1.12. More importantly,
we obtained $p=0.50\pm0.03$, that is just the value predicted by the theory of super-critical accretion disk (slim disk; \cite{44}; \cite{45}). 
We also obtained high inner disk temperature, $kT_{\rm in} = 2.54\pm0.34$, as is expected from large spectral hardening factors, which the slim disk model predicts. 
%The flux is $1.55\times10^{-12} {\rm erg}\ {\rm cm}^{-2}{\rm s}^{-1}$.

NGC 5204 X-1 is not an exception (see Table 1).
We also found a good fit with the $p$-free model for the cases of NGC 4559 X-7 and 
X-10. %The reduced chi-squared is 1.19 for X-7 and 0.93 for X-10. 
The $p$ values of $0.50 \pm 0.03$ (X-7) and $0.51 \pm 0.06$ (X-10)
and the inner disk temperatures of $1.84 \pm 0.16$ keV (X-7) and 
$3.16 \pm 0.71$ keV (X-10) are again consistent with the slim disk model.
%The fluxes are $7.96\times10^{-13} {\rm erg}\ {\rm cm}^{-2} {\rm s}^{-1}$ and 
%$9.14\times10^{-13} {\rm erg}\ {\rm cm}^{-2} {\rm s}^{-1}$ for X-7 and X-10, 
%respectively.

As for the cases of NGC 1313 X-2, 
fitting with the $p$-free model only also gave a good fit.
We obtained, $kT_{\rm in} = 2.00 \pm 0.34$ keV, and the $p$-value 
is $0.50 \pm 0.05$.

%and $8.05\times10^{-13} {\rm erg}\ {\rm cm}^{-2}{\rm s}^{-1}$ for X-2.

\subsubsection{Interpretation of $p$-free model}
It is clear from the results of the fitting with $p$-free model that the obtained {\it p} value is around 0.5, 
indicating super-critical accretion flow (slim disk) for all the data. 
Applying similar interpretation as in the case of fitting with DBB+PL, 
we obtained,$R_{\rm in}=23.7(\cos i)^{-1/2} (\kappa/1.7)^2$ km, the derived black hole mass 
of $2.67 \beta^{-1} (\cos i)^{-1/2} (\kappa/1.7)^2 M_{\odot}$ and $L_{\rm [0.3 - 10 keV]}/L_{\rm E} = 5.29 \beta (\cos i)^{-1/2} 
(\kappa/1.7)^{-2}$ for NGC 5204 X-1.

Caution is needed here, however, regarding the inner edge of the disks in the slim disk regimes and 
the plausible spectral hardening factor values ($\kappa$).  
Through the careful numerical integration, Watarai et al. (2000) found that
the inner edge of the disk is substantially smaller, 
$R_{\rm in}\sim R_{\rm S}$ (i.e., $\beta \sim 1/3$), in the slim disk regime, 
compared with that in sub-critical (standard disk) regime, 
$R_{\rm in}\sim 3 R_{\rm S}$ {\bf ($\beta \sim 1$)},
even though a central black hole is not rotating.  
This is because enhanced mass accretion flow with high density finally fills up 
an empty zone inside $3R_{\rm S}$, which exists in the sub-critical accretion regimes.  
We then expect significant blackbody emission from inside $3 R_{\rm S}$.
It is thus preferable to set $\beta \sim 1/3$, instead of $\beta = 1$.
However, radiation from the vicinity of black holes should be attenuated
because of intense gravitational redshifts and transverse Doppler effect, and since the effective temperature corrected for the relativistic effect has a peak at around $2 R_{\rm S}$, 
$\beta \sim 2/3$ will be a more reasonable estimate (\cite{20}). 

As for the spectral hardening factor, 
Shimura and Takahara (1995) and Zampieri et al. (2001) found $\kappa\approx1.7$  for 
stellar-mass black holes,  and it increases with 
increasing $\dot{M}$.  In the slim disk, 
the disk temperature gets higher and the density becomes smaller as $\dot{M}$ increases, therefore
the electron scattering becomes more dominant and the spectral hardening factor becomes much higher
 than $\sim1.7$ (\cite{20}). 
Watarai and Mineshige (2003) found from the comparison between the limit-cycle instability theory (\cite{18}) and the observation of GRS1915+105 that the spectral hardening 
factor close to the Eddington luminosity should be about 3.

The mass estimated from the $p$-free models is proportional to $\kappa^2$, 
so if we take a larger $\kappa$, say, 3, 
the inferred mass will be tripled compared to the case of standard disks ($\kappa \sim 1.7$).
Consequently, adoping $\beta \sim 2/3 $, our mass estimate is $12.5 (\cos i)^{-1/2}  (\kappa/3.0)^2 M_{\odot}$ and 
the Eddington ratio is  $L_{\rm [0.3 - 10 keV]}/L_{\rm E} \sim 1.1 (\cos i)^{-1/2}  (\kappa/3.0)^{-2}$ for NGC 5204 X-1.
In the same manner, we find smaller black hole masses and larger Eddington ratios for other sources (see Table 2). 
%we find $R_{\rm in}=65.7(\cos i)^{-1/2}$ km, the derived black hole mass of $7.4 \beta^{-1} (\cos i)^{-1/2}M_{\odot}$ and $L_{\rm [0.2 - 10 keV]}/L_{\rm E} = 4.03 \beta (\cos i)^{-1/2}$ for NGC 4559 X-7; and $R_{\rm in}=23.7(\cos i)^{-1/2}$ km, the derived black hole mass of $2.67 \beta^{-1} (\cos i)^{-1/2}M_{\odot}$ and $L_{\rm [0.3 - 10 keV]}/L_{\rm E} = 1.28\times10 ^{1} \beta (\cos i)^{-1/2}$ for NGC 4559 X-10,
%and $R_{\rm in}=12.8(\cos i)^{-1/2}$ km, the derived black hole mass of $1.44 \beta^{-1} (\cos i)^{-1/2}M_{\odot}$ and $L_{\rm bol}/L_{\rm E} = 2.1\times10^{1} \beta (\cos i)^{-1/2}$ for Holmberg II X-1; 
The derived black hole mass is well within the stellar mass black hole regime. 

%\tiny
%%%%%%%%%%%%%%%%%%%%%%%%%%%%%%%%%%%%%%%%
\begin{center}
\begin{longtable}{lllll}
  \caption{Fitting results with DBB+PL and $p$-free model. $N_{\rm H}$ is an absorption column external to our Galaxy in unit of $10^{21}$ atom cm$^{-2}$. A Galactic absorption column is assumed as in the text. $kT_{\rm in}$ is the inner disk temperature in keV. The normalization for DBB and p-free is defined by $[(R_{\rm in}/{\rm km})/(D/10{\rm kpc})]^{2}\cos{i}$, where $R_{\rm in}$ is the inner disk radius, $D$ the distance to the source, and $i$ the angle of the disk. The unit of power-law normalization is photon/s/cm2/keV at 1 keV. $f_{\rm x}$ is the observed flux in units of erg cm$^{-2}$ s$^{-1}$ ($0.3 - 10$ keV) that depends on the model.
}\label{tab:first}
 \hline \hline
  & & & & \\
  Parameter & NGC 5204 X-1 & NGC 4559 X-7 & NGC 4559 X-10 & NGC 1313 X-2  \\
  & & & & \\
%\endfirsthead
  \hline\hline
\endhead
 \hline
\endfoot
  \hline
\endlastfoot
 & & & & \\
 DBB+PL & & & & \\
 & & & & \\
 $N_{\rm H}$ & $0.51\pm 0.11$ & $1.58\pm 0.19$ & $0.98\pm 0.22$ & $1.98\pm 0.27$ \\
 $kT_{\rm in}$ & $0.25\pm 0.03$ & $0.17\pm 0.09$ & $0.48\pm 0.25$ & $0.27\pm 0.04$ \\
 Norm. & $3.7\pm 2.6$ & $46.9\pm 30.3$ & $0.04\pm 0.11$ & $4.5\pm 4.0$ \\
 $\Gamma$ & 1.92 $\pm 0.06$ & 2.17 $\pm 0.05$ & 1.94 $\pm 0.13$ & 2.01 $\pm 0.13$ \\
 Norm. & ($3.02\pm 0.48$)$\times10^{-4}$ & ($2.10\pm 0.24$)$\times10^{-4}$ & ($2.01\pm 0.83$)$\times10^{-4}$ & ($2.02\pm 0.73$)$\times10^{-4}$ \\
 $f_{\rm x}$ & $1.66\times10^{-12}$ & $8.36\times10^{-13}$ & $9.42\times10^{-13}$ & $8.63\times10^{-13}$ \\
 $\chi^2/ \rm {dof}$ & 1.05 & 0.96 & 0.93 & 0.98 \\
   & & & & \\
  \hline
 & & & & \\
 p-free & & & & \\
 & & & & \\
 $N_{\rm H}$ & $0.34\pm 0.01$ & $0.60\pm 0.06$ & $0.91\pm 0.08$ & $1.42\pm 0.14$ \\
 $kT_{\rm in}$ & $2.54\pm 0.34$ & $1.84\pm 0.16$ & $3.16\pm 0.71$ & $2.03\pm 0.35$ \\
 $p$ & $0.50 \pm 0.03$ & $0.50 \pm 0.03$ & $0.51 \pm 0.06$ & $0.50 \pm 0.05$ \\
 Norm. & ($5.04\pm 6.11$)$\times10^{-4}$ & ($1.04\pm 0.85$)$\times10^{-3}$ & ($1.51\pm 3.13$)$\times10^{-4}$ & ($8.13\pm 12.9$)$\times10^{-4}$ \\
 $f_{\rm x}$ & $1.55\times10^{-12}$ & $7.96\times10^{-13}$ & $9.14\times10^{-13}$ & $8.08\times10^{-13}$ \\ 
 $\chi^2/ \rm {dof}$ & 1.12 & 1.19 & 0.93 & 1.15 \\
   & & & & \\
   \hline \hline 
\end{longtable}
\end{center}

%\normalsize

\subsection{Fitting Results with Kawaguchi's model}
Finally, we present the results of the data fitting with the slim disk model developed by Kawaguchi (2003) (see Foschini et al. 2005 for the case of ULX M33 X-8). He developed several models, but we chose the model in which the effects of electron scattering and relativistic correction are included. 
Calculation was made and fitting was tried for several different values of the viscous parameter, $\alpha$.
We found  $\alpha= 0.1$ gives most satisfactory fits, and will show only the results with  $\alpha= 0.1$ in this paper.
Source distance was fixed, and the face-on geometry  assumed.  Model parameters  are only mass and mass accretion rate
(in the unit of $L_{\rm Edd}/c^2$).
We summarize the fitting results using this model in Table 3.

The best fit for NGC 5204 X-1 was obtained with reduced chi-squared of 1.18,  black hole mass of $18 M_{\odot}$ and mass accretion rate of $27 \dot{M}_{\rm Edd}$,
where $\dot{M}_{\rm Edd} \equiv L_{\rm Edd}/c^2$
\footnote{Here,
$16 L_{\rm Edd}/c^2$ is the critical accretion rate, which would
produce Eddington luminosity in the case of a classical
radiative efficiency ($1/16$).}.
All the analyzed data of ULXs can be fitted with Kawaguchi's model and 
the derived black hole masses are less than 100 $M_\odot$ (see Table 3).
This is not so surprising for the data showing the slim disk signatures, 
because the basic spectral shapes of his model are similar to
those of the $p$-free model with $p\sim 0.5$
(as long as $\dot M\gg L_{\rm E}/c^2$).

%%%%%%%%%%%%%%%%%%%%%%%%%%%%%%%%%%%%%%%%%%%%%%%%%%%

\begin{center}
\begin{longtable}{lllll}
  \caption{The derived black hole mass value from DBB+PL model and $p$-free model in $M_{\odot}$ and the ratio between luminosity at 0.3 -- 10 keV and Eddington luminosity. We assumed $\beta \sim 1$ for DBB+PL model and  {\bf $\beta \sim 2/3$} for $p$-free model (see text).
The spectral hardening factor, $\kappa$,  is assumed to be 1.7 for DBB model, and considered to be $\approx3 $ 
for slim disk.}\label{tab:second}
  \hline\hline
   & NGC5204 X-1 & NGC4559 X-7 & NGC4559 X-10 & NGC1313 X-2 \\
\hline\hline
   & & & & \\
  DBB+PL & & & & \\
\endhead
  \hline
\endfoot
  \hline
\endlastfoot
 $M(\cos i)^{1/2}$ & $287$ & $889$ & $118$ & $136$ \\
 $L(\cos i)^{1/2}/L_{\rm E}$ & $0.05$ & $0.03$ & $0.3$ & $0.03$ \\
 & & & & \\
 \hline
  & & & & \\
 $p$-free & & & & \\
 $M(\cos i)^{1/2}$ & $12.5(\frac{\kappa}{3})^2$ & $34.6(\frac{\kappa}{3})^2$ & $12.5(\frac{\kappa}{3})^2$ & $11.2(\frac{\kappa}{3})^2$ \\
$L(\cos i)^{1/2}/L_{\rm E}$ & $1.1(\frac{\kappa}{3})^{-2}$ & $0.9(\frac{\kappa}{3})^{-2}$ & $2.8(\frac{\kappa}{3})^{-2}$ & $0.4(\frac{\kappa}{3})^{-2}$ \\ 
 & & & & \\
  \hline \hline 
\end{longtable}
\end{center}

%%%%%%%%%%%%%%%%%%%%%%%%%%%%%%%%%%%%%%%%%%%%%%%%%%%
\begin{center}
\begin{longtable}{lllll}
  \caption{Fitting results with slim disk model of Kawaguchi (2003). $N_{\rm H}$ is an absorption column external to our Galaxy in unit of $10^{21}$ atom cm$^{-2}$. A Galactic absorption column is assumed as in the text. $M$ is mass in $M_\odot$. $\dot{M}$ is mass accretion rate ($=L_{\rm Edd}/c^{2}$). $f_{\rm x}$ is the observed flux in units of $10^{-12}$ erg cm$^{-2}$ s$^{-1}$ (0.3 -- 10 keV) that depends on the model.   }\label{tab:second}
  \hline\hline

  Fit Par. & NGC5204 X-1  & NGC4559 X-7 & NGC4559 X-10 & NGC1313 X-2 \\

\hline\hline
%  $\alpha=0.1$ & & & & \\
\endhead
  \hline
\endfoot
  \hline
\endlastfoot
 & & & & \\
 $N_{\rm H}$ & $0.09\pm 0.03$ & $0.45\pm 0.03$ & $0.69\pm 0.07$ & $1.23\pm 0.07$ \\
 $M$ & $18^{+2}_{-2}$ & $57^{+4}_{-4}$ & $24^{+9}_{-7}$ & $12^{+1}_{-1}$ \\
 $\dot{M}$ & $27^{+5}_{-4}$ & $18^{+2}_{-1}$ & $100^{+70}_{-40}$ & $14^{+2}_{-1}$ \\
 $f_{\rm x}$ & $1.6$ & $0.79$ & $0.91$ & $0.8$ \\
 $L_{\rm [0.3 - 10 keV]}/L_{\rm E}$ & 0.81 & 0.52 & 1.42 & 0.36 \\
 $\chi^2/ \rm {dof}$ & 1.18 & 1.20 & 0.93 & 1.17\\
 & & & & \\
 \hline \hline
\end{longtable}
\end{center}

%%%%%%%%%%%%%%%%%%%%%%%%%%%%%%%%%%%%%%%%%%%%%%%%%%%

\section{Discussion}

\subsection{Accretion Disk Models and Black Hole Mass}

In the present study, we re-examined the XMM-Newton data of
ULXs through the conventional and more sophisticated spectral models
constructed based on the theoretical study of super-critical accretion flow.
We would like to comment on the results from fitting with DBB+PL and $p$-free model first. It is obvious from the fitting results in other papers, 
e.g.Roberts et al. (2005), Cropper et al. (2004), Miller et al. (2003,2004), 
that the DBB+PL model provides a good fit to several ULXs data, 
and the 'cool' disk temperature obtained from the fitting suggests an existence of IMBHs inside the ULXs. 
However, we should note that this is only one option, and we tried another option; namely, the $p$-free model.  Remarkably, 
the $p$-free model also gives good fit to the same data. 
Moreover, the obtained {\it p} values, which are $\sim 0.5$, 
indicating the super-critical accretion (slim disk) model, 
in stead of standard disk (DBB) model where, $p = 0.75$.
The derived black hole masses and the Eddington ratios also support this conclusion.

Why can the two distinct models give similarly good fits to the data?
A trick lies in the spectral slope of the slim disk model.
For a power-law temperature profile, $T \propto r^{-p}$, the entire disk spectra, 
which is the summation of the blackbody emission spectra from various radii, 
give power-law spectral slope in the intermediate energy range (below $h\nu < kT_{\rm in}$) with a spectral slope of $F_\nu \propto \nu^{3-(2/p)}$ 
(see, e.g., chapter 3 of Kato et al. 1998).  
Then, a photon index is $\Gamma = (2/p) - 2$. 
That is, for a slim disk with $p=1/2$, we have $\Gamma = 2$.  
(Note that for a standard disk with $p=3/4$ we get $\Gamma = 2/3$.)  
This is a striking coincidence that both of the PL model with $\Gamma = 2$ 
and the slim disk model give the same power-law spectra in the moderate energy range. 
To see this more explicitly, we show in Fig. 2 the two fitting models with fitting parameters obtained for NGC 5204 X-1.
We see that both DBB+PL model and slim disk model produce power-law decline with a power-law index $\sim -1$ (or photon index, $\Gamma\sim2$) 
and are identical at $0.5\sim5$ keV energy range.
%which answer the question of why we could obtain a good fit in both DBB+PL and $p$-free models, $F\propto\nu^{-1}$, while DBB model alone which represents a standard disk interpretation will produce $F\propto\nu^{1/3}$.  

In the present paper, it was hardly possible
from model fitting to judge if the DBB+power-law model or the slim disk modesl is preferred,
being limited by photon statistics.
It is, however,  in principle possible to distinguish the 
two models by measuring  their spectra very accurately.
So far, XMM-Newton has been able to do this only for M82 X-1,
the brightest ULX (\cite{33}). Okajima et al. (2006)
found that the energy spectrum of M82 X-1 above $\sim$ 3 keV has a curvature just between
the power-law and DBB model, and is represented pretty well with $p$-free model of $p=0.61$.
This is considered to be strong evidence of slim disk in ULXs.
Next generation hard X-ray satellites  with higher sensitivity and imaging capability 
are expected to precisely measure the energy spectra of most ULXs, so that black hole parameters
are consrained by applying precise slim disk model spectra.

\subsection{Luminosity-Temperature diagram}

Then, which is more likely the case, 
sub-critical accretion onto intermediate-mass black holes or 
super-critical accretion onto stellar-mass black holes?

We pointed out a serious problem inherited from the DBB+PL model.
The spectral fitting with DBB+PL model shows that the power-law component predominates in all energy regions (Fig. 2). 
In other words, disk contribution is very small compared to the power-law contribution 
and therefore the black hole masses derived above might not be so reliable, 
because the standard disk relations cannot be used here.
%and also because we do not have a good constraint when we derived those values.

On the other hand, a good fit with $p$-free model alone implies that the disk contributes in all energy region, 
therefore, the black hole masses derived from the fitting with $p$-free model can be reliable. 
For these reasons, we can safely conclude that the latter model is more likely the case to explain the physics of ULXs.

We summarize our results in Fig. 3. 
This is the temperature-luminosity diagram (cf. \cite{25};
cf. \cite{30}) with the lines of constant black-hole masses. 
We plot our ULXs fitting results with DBB+PL and $p$-free model 
together with several best-studied stellar-mass black hole candidates (BHCs).  
We can see that the low temperatures obtained from the fitting using the DBB+PL model are shifted to higher temperatures when fitted with the $p$-free model. 
This implies that the black hole mass values in the range of IMBHs obtained from the fitting with the DBB+PL model can be shifted to the stellar mass range by the $p$-free model; 
that is, still there is no evidence of IMBHs so far, at least in four ULXs investigated in this paper, 
due to the minor disk contribution in DBB+PL model as mentioned before.

%%%%%%%%%%%%%%%%%%%%%%%%%%%%%%%%%%%%%%%%%%%%%%%%%%%%
\begin{figure*}
  \begin{center}
%    \FigureFile(120mm,120mm){logcomparexerrbar.ps}
    %%% \FigureFile(width,height){filename}
\centerline{\epsfig{file=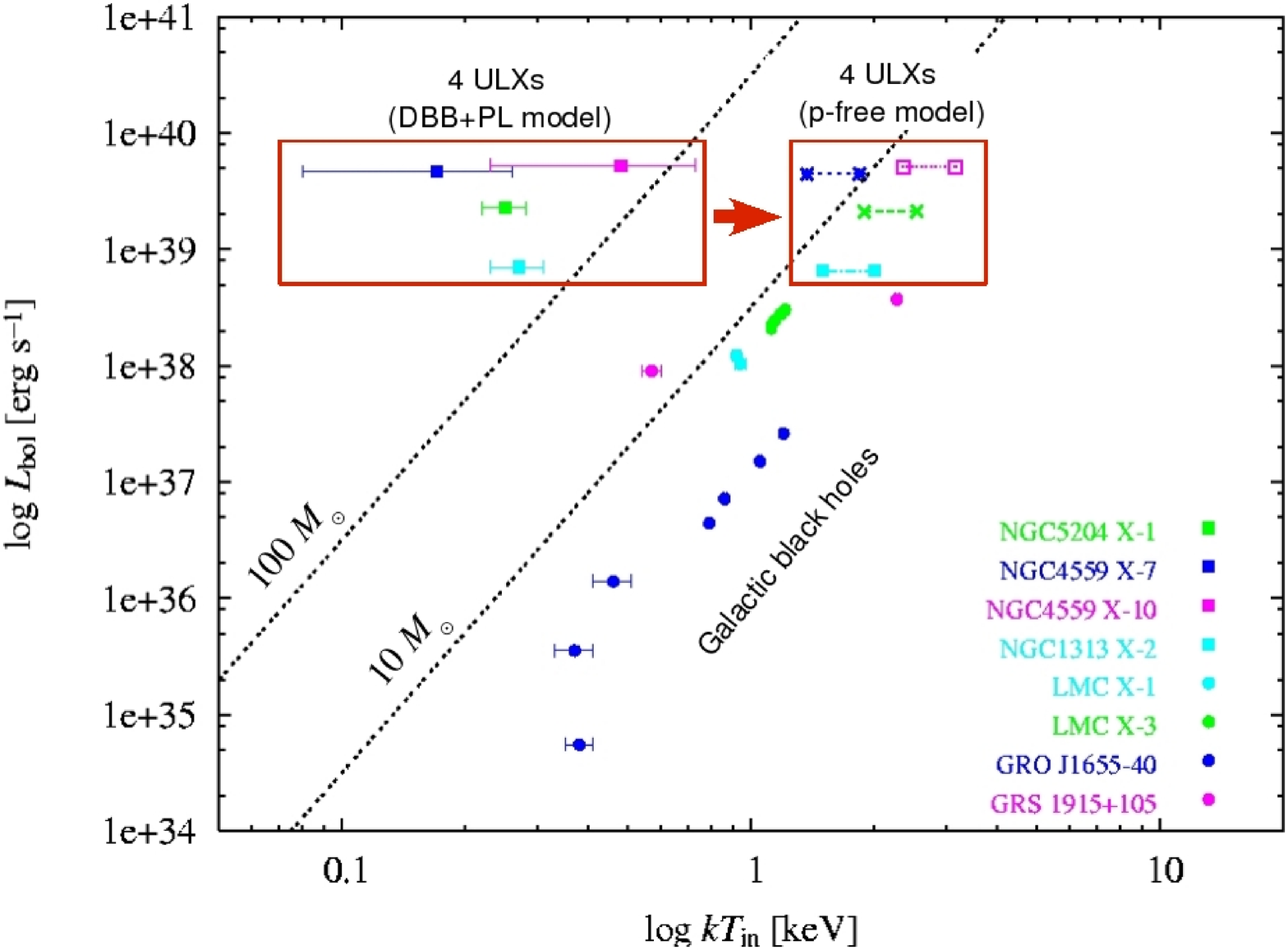,width=14.0cm,angle=0}}
  \end{center}
  \caption{Relation between the luminosity at 0.3 -- 10 keV ($L_{\rm [0.3 - 10 keV]}$) 
and $T_{\rm in}$ for ULXs and stellar mass black hole candidates (BHCs).
%Black dotted line shows the line for $10M_\odot$ and green dashed line represents the line for $100M_\odot$. 
BHCs references: Ebisawa 1991 (LMC X-1), Treves et al. 1988 (LMC X-3), 
Mendez et al.1998 (GRO J1655-40), and Belloni et al. 1997 (GRS 1915+105). Note that in this figure $L_{\rm bol}$ means the
0.3 -- 10  keV luminosity  for ULXs and the bolometric luminosity for Galactic black hole candidates, both assuming the face-on
geometry. 
The ranges of $T_{\rm in}$ in the $p$-free model box represent the values via DBB model (lowest end) and 
$p$-free model (highest end).}\label{Figure:5}
\end{figure*}

%%%%%%%%%%%%%%%%%%%%%%%%%%%%%%%%%%%%%%%%%%%%%%%%%%%%

It then naturally follows that
all the fitting results which gave low disk temperatures and photon indices of $\Gamma \sim 2$ should be re-examined, 
since those data are very likely to be fitted with the slim disk ($p$-free) model equally well.  
Obviously, the latter model gives higher disk temperatures and hence lower black hole masses.  
It is, in this respect, interesting to note the statistical study of ULXs by Winter et al. (2006) who examined distribution of photon indices and found a concentration at $\Gamma \sim 2$ for the high-state ULXs, 
those ULXs whose spectra can well be fitted with the combined DBB+PL model.  
Some of them may be in the slim disk regimes, rather than in the high state.

\section{Conclusion}

We have investigated the XMM-Newton EPIC spectra of the four ULXs
which were claimed to be strong candidates of IMBHs by several authors.
We found that these spectra are successfully fitted with slim disk models, and the derived masses are
$\sim $11 to 60 $M_\odot$ depending on different assumptions.  We have not seen the evidence of
the ``intermediate mass'' black holes having several hundreds of $M_\odot$.
When the mass accretion rate is close to or exceeding the critical rates, the slim disk is theoretically predicted.
We suggest that ULXs are stellar mass black holes, at most several tens of $M_\odot$ under super-critical accretion rates
shining at super-Eddington luminosities.

\bigskip
\bigskip
We gratefully thank Marek Abramowicz for valuable comments and suggestions. This work was supported in part by the Grants-in-Aid of the Ministry
of Education, Science, Culture, and Sport  (14079205, 16340057 S.M.), by the Grant-in-Aid for
the 21st Century COE ``Center for Diversity and Universality in Physics'' from the Ministry
of Education, Culture, Sports, Science and Technology (MEXT) of Japan.
K.V. gratefully  thank Ken-ya Watarai for the discussions at some beginnings of this work and also MEXT scholarship. 
T.K. thanks the financial supports from JSPS Postdoctoral Fellowship (01879). 

%%%
% See the manual for the detail.
%%%

\end{document}